\documentclass[11pt,a4paper]{article}
\usepackage{a4wide}

\usepackage[dvipdfmx,hiresbb]{graphicx} 
\usepackage{mediabb}                    

\usepackage{amsmath,amssymb}
\usepackage{color}
\usepackage[bookmarks,bookmarksnumbered]{hyperref}
\usepackage{cellspace}
\usepackage{tikz}
\usetikzlibrary{fadings}
\usetikzlibrary{patterns}
\usetikzlibrary{shadows.blur}
\usetikzlibrary{shapes}

\usepackage{amsmath,amssymb}
\usepackage{mathtools}
\usepackage{cite}
\usepackage{authblk}
\usepackage{mathrsfs}
\usepackage{color}
\usepackage{bm}

\title{Swampland Bounds on Magnetized Extra Dimensions}
\author[1,2]{Yuta Hamada\footnote{yhamada@post.kek.jp}}
\author[3]{Maki Takeuchi\footnote{maki\_t@yamaguchi-u.ac.jp}}
\affil[1]{\small \textit{Theory Center, IPNS, High Energy Accelerator Research Organization (KEK), 1-1 Oho, Tsukuba, Ibaraki 305-0801, Japan}}
\affil[2]{\small \textit{Graduate Institute for Advanced Studies, SOKENDAI, 1-1 Oho, Tsukuba, Ibaraki 305-0801, Japan}}
\affil[3]{\small \textit{Graduate School of Sciences and Technology for Innovation, Yamaguchi University, Yamaguchi-shi, Yamaguchi 753-8512, Japan}}

\date{}

\begin{document}

\begin{flushright}
 KEK-TH-2628   
\end{flushright}

{\let\newpage=\relax \maketitle}

\begin{abstract}
We consider a six-dimensional $U(1)$ gauge theory compactified on two-dimensional manifolds.
The number of chiral fermions is determined by the flux quantization number on the two-dimensional compact manifolds. 
Using the Swampland Conjectures, we find constraints among the parameters of the theory: the flux quantization number, the compactification scale, and the string scale. Specifically, the Weak Gravity Conjecture and the Trans-Planckian Censorship Conjecture give non-trivial bounds. 
\end{abstract}

\newpage  

\tableofcontents   

\newpage

\section{Introduction}

Higher dimensional theories with extra dimensions, such as string theory, are prominent candidates for theories beyond the Standard Model. The compactification of extra dimensions is crucial in obtaining a four-dimensional low-energy theory. 
Extra-dimensional models incorporating phenomenologically favored features such as three generations and flavor structures have been actively explored. 

If the extra dimensions are two-dimensional,  candidate manifolds for the compactifications are topologically spheres $S^2$~\cite{Dolan:2020sjq}, torus $T^2$~\cite{Cremades:2004wa}, and compact hyperbolic spaces $H^2/\Gamma$~\cite{Arefeva:1985yn,Orlando:2010kx}. Particularly, in the presence of magnetic fluxes, it is possible to obtain a four-dimensional chiral theory (see e.g. \cite{Kariyazono:2019ehj,Ohki:2020bpo,Kikuchi:2020frp,Kikuchi:2023awe,Hirose:2024vvx} for recent studies).
In these models, zero mode degeneracy occurs, and the degeneracy number is equal to the number of generations. The number of generations is computed from the Atiyah-Singer index theorem~\cite{Atiyah:1963zz}, which states that the number of chiral zero modes for the Dirac operator is equal to the topological invariant. 
In the case of two dimensions, the flux quantization number corresponds to the topological invariant.
Consequently, extra-dimensional models stand as compelling theories beyond the Standard Model. However, although these models can explain the multiple generations, it is difficult to single out three generations. This is because the flux quantization number is an arbitrary integer. 
As a constraint from the experiment, the compactification scale is suggested to be ${\mathcal{O}}(10)$ TeV or higher. This is because Kaluza-Klein particles have not yet been discovered in the LHC experiments.

The Swampland conditions~\cite{Vafa:2005ui} provide constraints on the parameter space of effective field theories. 
See Refs.~\cite{Palti:2019,vanBeest:2021lhn,Grana:2021zvf,Agmon:2022thq} for the reviews. 
These are mainly motivated by the consideration of black hole physics, and examples of string Landscape. 
The Swampland Conjecture provides a potential avenue to constrain the parameters of the extra dimension. 
Among various Swampland conjectures, we utilize the Weak Gravity Conjecture (WGC)~\cite{Arkani-Hamed:2006emk} (see \cite{Harlow:2022ich} for a review) and the Trans-Planckian Censorship Conjecture (TCC)~\cite{Bedroya:2019snp,Bedroya:2019tba} to constrain the parameter space of the magnetized extra dimensions.\footnote{See e.g. \cite{Cheung:2016wjt,Hamada:2018dde,Noumi:2022ybv} for a bottom-up explanation of the WGC, and \cite{Arkani-Hamed:2006emk,Heidenreich:2016aqi,Montero:2016tif,Lee:2018urn} for tests of the WGC in string theory.} 
An idea is that, in the presence of the flux, the vacuum is unstable under the nucleation of the membrane~\cite{Brown:1988kg}. The presence of the dynamical membrane with the finite tension is guaranteed by the WGC~\cite{Ooguri:2016pdq}.
The metastability of the vacuum against the membrane creation places a non-trivial constraint on the model parameters.
Moreover, the existence of the vacuum with flux number three indicates that there exists the de Sitter vacua with the flux numbers four, five, six, and so forth.
Even though we do not live in these vacua, the TCC sets the upper bound for the lifetime of de Sitter spacetime. This provides further constraints on the parameter space.

This paper is organized as follows. In Section \ref{models}, we review the models of two-dimensional compact manifolds with magnetic flux. In Section \ref{Membranecreation}, we review the membrane pair nucleation in the flux vacua. In Section \ref{main}, we constrain the parameters by some of the Swampland Conjectures and the stability condition. Section \ref{Conclusion} is devoted to the discussion and conclusion.

\section{Magnetized extra dimensional models}
\label{models}
We consider the $U(1)$ gauge theory on two-dimensional compact manifolds (a two-sphere $S^2$, a 2d torus $T^2$, and a compact hyperbolic manifold $H^2/\Gamma$~\cite{Arefeva:1985yn,Orlando:2010kx}) with homogeneous magnetic flux. In these manifolds, the gauge field strength satisfies the flux quantization condition~\cite{Witten:1984dg}
\begin{align}
    \frac{1}{2\pi}\int_{{\cal{M}}^2}F_2=M, \label{fluxquantized}
\end{align}
where $F_2$ is a 2-form field strength of the $U(1)$ gauge field, $M$ is a flux quanta, and ${\cal{M}}^2$ is a 2d compact manifold.

The Atiyah-Singer index theorem for a 2d compact manifold ${\cal{M}}^2$ with magnetic flux is known as~\cite{Witten:1984dg}
\begin{align}
n_{+}-n_{-}=\frac{1}{2\pi}\int_{{\cal{M}}^2}F_2,
\label{Ind2d}
\end{align}
where $n_{\pm}$ are the numbers of $\pm$ left and right chiral zero modes for the Dirac operator.
Substituting \eqref{fluxquantized} into \eqref{Ind2d}, we get
\begin{align}
    n_{+}-n_{-}=\frac{1}{2\pi}\int_{{\cal{M}}^2}F_2=M.
    \label{Ind2d2}
\end{align}
The index theorem \eqref{Ind2d2} indicates that the generation number of the Standard Model is given by $M$.

\section{Membrane nucleation}
\label{Membranecreation}

\subsection{Set up}
Taking a Hodge dual of the field strength $F_2$ on the magnetized compact 2d manifold, the system is equivalent to the one with a 4-form flux $F_4$ in four-dimension:
\begin{align}
    \ast_6 F_2=F_4.
\end{align}
Thus, the field strength $F_2$ in two-dimension can be read off as a 4-form antisymmetric tensor field $F_4$ in four-dimension.
In \cite{Brown:1988kg}, the production rate of membranes in the presence of the flux of the 3-form gauge field $A_{\mu_1\mu_2\mu_3}$ was discussed, where the membranes are charged under $A_{\mu_1\mu_2\mu_3}$. 
This can be viewed as a higher form generalization of the Schwinger effect~\cite{Schwinger:1951nm}. Note that the Schwinger effect is a phenomenon in which electron-positron pairs are spontaneously created in the presence of an electric field.
The change in the two-dimensional field strength $F_2$ can be applied to the picture of membrane formation by the four-dimensional 4-form antisymmetric tensor field $F_4$.

The bubble nucleation is described as an instanton tunneling process~\cite{PhysRevD.15.2929,PhysRevD.16.1762,PhysRevD.21.3305} in the semiclassical approximation. Instantons are solutions of the classical Euclidean equations of motion. Phase transitions occur through the spontaneous appearance in the metastable phase of closed domain walls or bubbles enclosing a true vacuum region. These bubbles are initially formed in a quiescent state, then evolve classically, expand rapidly, and merge with other bubbles.

\subsection{Membrane creation}
In this subsection, we review a Brown-Teitelboim picture of membrane nucleation\cite{Brown:1988kg}. 
Let $x^{\mu}=z^{\mu}(\xi)$ denote the three-dimensional worldvolume of the membrane in 4d spacetime as a function of worldvolume coordinates $\xi^{a}\,\,(a=1,2,3)$. 
The Euclidean action is
\begin{align}
    S_{\rm{E}}&=m\int d^3 \xi \sqrt{^3 g}+\frac{e}{3!}\int d^3 \xi A_{\mu_1 \cdots \mu_3}\left[\frac{\partial z^{\mu_1}}{\partial \xi^{a_1}}\cdots\frac{\partial z^{\mu_3}}{\partial \xi^{a_3}} \right]\varepsilon^{a_1\cdots a_3}-\frac{1}{2\cdot 4!}\int d^4x\sqrt{ g} F_{\mu_1 \cdots \mu_4}F^{\mu_1 \cdots \mu_4}
     \notag \\
   &\qquad  +\frac{1}{3!}\int d^4 x \partial_{\mu_1}\left[\sqrt{ g} F^{\mu_1 \cdots \mu_4}A_{\mu_2 \cdots \mu_4}\right]+S^{\rm{grav}}_{\rm{E}}(\lambda),
   \label{action1}
\end{align}
where the gravitational action $S_E^{\rm{grav}}(\lambda)$ is
\begin{align}
  S^{\rm{grav}}_{\rm{E}}(\lambda)=-\frac{1}{16\pi G}\int d^4x  \sqrt{ g}(R-2\lambda)+\frac{1}{8\pi G} \oint d^3 x  \sqrt{ h} K .
  \label{gaction}
\end{align}
Here, $m$ is the mass per unit volume of the membrane, $^3 g_{ab}=g_{\mu \nu}{z^{\mu}}_{,a}{z^{\nu}}_{,b}$ is the induced metric, $e$ is the coupling constant between the membrane and the antisymmetric tensor field, $\lambda$ is the bare cosmological constant.
The Gibbons-Hawking~\cite{PhysRevD.15.2752} surface term in \eqref{gaction} is an integral over all boundaries of the 4d spacetime with $h$ being the induced metric and $K$ being the trace of the extrinsic curvature. This term is included to ensure that the gravitational action has well-defined functional derivatives with respect to variations in the normal derivative of the metric at the boundary. 
The total derivative term in \eqref{action1} is the topological invariant discussed in~\cite{Aurilia:1980xj}, which must be included to ensure that the action has well-defined functional derivative with respect to $A$. 

\begin{figure}[h]
\centering
\includegraphics[width=0.6\textwidth]{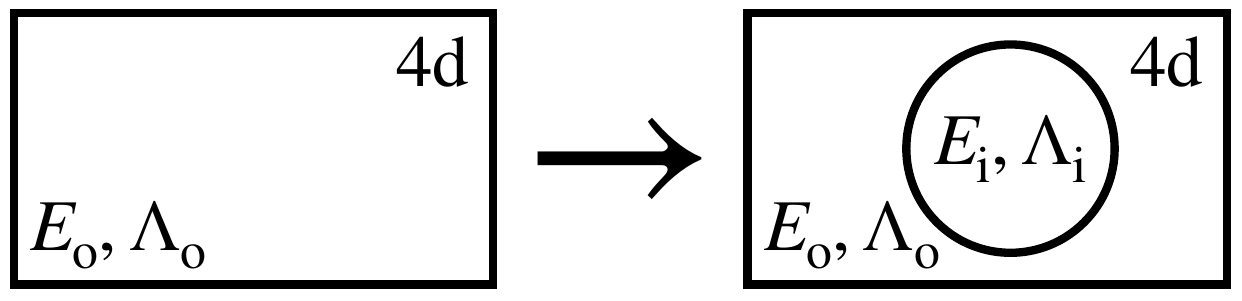}
    \caption{Schematic picture of the membrane nucleation.}
    \label{Fig_membrane}
\end{figure}

The 4-form $F_4$ tensor can be written as
\begin{align}
    F^{\mu_1 \cdots \mu_4}=\frac{E}{\sqrt{ g}}\varepsilon^{\mu_1\cdots \mu_4},
\end{align}
for a scalar field $E$.
The equation of motion from \eqref{action1} for the 3-form field $A_{\mu_1\mu_2\mu_3}$ is 
\begin{align}(\partial_{\mu_1}E)\varepsilon^{\mu_1\cdots \mu_4}=
    -e\int d^3 \xi \,\delta^4(x-z(\xi))\left[\frac{\partial z^{\mu_2}}{\partial \xi^{a_1}}\cdots\frac{\partial z^{\mu_4}}{\partial \xi^{a_3}} \right]\varepsilon^{a_1\cdots a_3}.
    \label{EOM1}
\end{align}
Substituting \eqref{EOM1} into \eqref{action1}, we get
\begin{align}
    S_{\rm{E}}&=m\int d^3 \xi \sqrt{^3 g}+\frac{1}{2\cdot 4!}\int d^4x\sqrt{ g} F_{\mu_1 \cdots \mu_4}F^{\mu_1 \cdots \mu_4}
     +S^{\rm{grav}}_{\rm{E}}(\lambda).
   \label{action2}
\end{align}
Eq.~\eqref{EOM1} shows that on either side of the membrane as the delta function vanishes.
Moreover, the values of $E$ differ by $|e|$. 
Consequently, the membrane divides the space into two regions (outside and inside) having different energy density contributed by the antisymmetric tensor field.
The outside and inside field values are related by
\begin{align}
    E_{\rm{i}}=E_{\rm{o}}-e,
\label{eq:E_difference}\end{align}
where the subscripts refer to outside and inside, and we assume $E_{\rm{i}}<E_{\rm{o}}$.
From the equation of motion of $g_{\mu\nu}$ and \eqref{eq:E_difference}, the outside and inside cosmological constants are 
\begin{align}
    &\Lambda_{\rm{o}}=\lambda+4\pi G(E_{\rm{o}})^2,\quad
    &\Lambda_{\rm{i}}=\Lambda_{\rm{o}}-8\pi G\left(eE_{\rm{o}}-\frac{1}{2}e^2\right)\label{Lambdainside}.
\end{align}
The schematic picture of the membrane nucleation is shown in Figure \ref{Fig_membrane}.

In this system, the membrane plays the role of a domain wall, and the energy density of the antisymmetric tensor field replaces the energy in the two vacuum states. The energy density is reflected in the spacetime geometry as a contribution to the cosmological constant. Consider a spacetime that initially contains a constant field $E=E_{\rm{o}}$, the cosmological constant $\Lambda=\Lambda_{\rm{o}}$, and no membranes. Once a membrane is created, it will encompass a region with new ``inside" field values $E_{\rm{i}},\,\Lambda_{\rm{i}}$, while the ``outside" values remain $E_{\rm{o}},\,\Lambda_{\rm{o}}$ as shown in Figure \ref{Fig_membrane}.

Ref.~\cite{Brown:1988kg} found that membrane nucleation is not always possible. The possibility of membrane nucleation depends on the values of $\sigma_{\rm{o}}$, $\sigma_{\rm{i}}$, sign $\Lambda_{\rm{o}}$, and sign $\Lambda_{\rm{i}}$. Here $\sigma_{\rm{o}}$ and $\sigma_{\rm{i}}$ are given by
\begin{align}
    &\sigma_{\rm{o}}={\rm{sign}}\left[\frac{1}{3}\left(eE_{\rm{o}}-\frac{e^2}{2}\right)-2\pi G m^2\right],
    &\sigma_{\rm{i}}={\rm{sign}}\left[\frac{1}{3}\left(eE_{\rm{o}}-\frac{e^2}{2}\right)+2\pi G m^2\right].\label{sigmainandout}
\end{align}
In Tab.~\ref{T1} we list the cases where membrane nucleation is possible(p) or not(n).

\begin{table}[h]
 \centering
 \begin{tabular}{c|c|c|c|c}
\hline
 & $\Lambda_{\rm{o}}>0,\,\sigma_{\rm{o}}=+1$ &
 $\Lambda_{\rm{o}}>0,\,\sigma_{\rm{o}}=-1$ &
 $\Lambda_{\rm{o}}\leq 0,\,\sigma_{\rm{o}}=+1$ &
 $\Lambda_{\rm{o}}\leq 0,\,\sigma_{\rm{o}}=-1$ \\ \hline
 $\Lambda_{\rm{i}}>0,\,\sigma_{\rm{i}}=+1$ &p& p&n&p\\ \hline
  $\Lambda_{\rm{i}}>0,\,\sigma_{\rm{i}}=-1$ & n&p&n&p\\ \hline
   $\Lambda_{\rm{i}}\leq0,\,\sigma_{\rm{i}}=+1$ & p&p&p&p\\ \hline
    $\Lambda_{\rm{i}}\leq0,\,\sigma_{\rm{i}}=-1$ &n &n&n&p\\ \hline
 \end{tabular}
 \caption{Membrane creations are possible or not. p means possible and n means not possible.}
 \label{T1}
 \end{table}

When the membrane nucleation is possible, the transition rate is obtained from the Euclidean action $S_{\rm{E}}$ as
\begin{align}
    \Gamma=Pe^{-B} \qquad B=S_{\rm{E}}[\rm{instanton}]-S_{\rm{E}}[\rm{background}].
    \label{bubble}
\end{align}
Here, $S_{\rm{E}}[\rm{instanton}]$ and $S_{\rm{E}}[\rm{background}]$ are the Euclidean actions evaluated at the instanton and background configurations.
$P$ is a prefactor that includes quantum fluctuations. 

The explicit form of the bounce action $B$ is given in~\cite{Brown:1988kg}: 
\begin{align}
    &B=mA_3(\bar{\rho})+\left\{\left[-\frac{2\Lambda_{\rm{i}}}{16\pi G}V_4(\bar{\rho},\sigma_{\rm{i}},\Lambda_{\rm{i}})-\frac{3\sigma_{\rm{i}}}{8\pi G}\left(\bar{\rho}^{-2}-\frac{2\Lambda_{\rm{i}}}{6}\right)^{\frac{1}{2}}A_3(\bar{\rho})\right]-({\rm{i}}\to {\rm{o}})\right\}, \label{B}
\end{align}
where $\bar{\rho}$ is the proper radius of the membrane, $A_3(\bar{\rho})$ is the area of the 3-dimensional Euclidean membrane, and $V_4$ is the 4-dimensional volume of the inside region. These are written as 
\begin{align}
    &\bar{\rho}=\left[\frac{2\Lambda_{\rm{o}}}{6}+\frac{1}{m^2}\left\{\frac{1}{3}\left(eE_{\rm{o}}-\frac{e^2}{2}\right)-2\pi G m^2\right\}^{2}\right]^{-\frac{1}{2}}, \label{eq:rhobar} \\
    &A_3(\bar{\rho})=\int d^3 \xi \sqrt{^3 g}=\frac{2\pi^2}{\Gamma(2)}\bar{\rho}^3, \\
    &V_4(\bar{\rho},\sigma_{\rm{\alpha}},\Lambda_{\rm{\alpha}}(>0))=\int_{\rm{inside}}d^4x 
    \sqrt{g}
    =\left(\frac{3}{|\Lambda_{\rm{\alpha}}|}\right)^2 \frac{A_3(\bar{\rho})}{\bar{\rho}^3}
    \left|\int_{1}^{\sigma_{\rm{\alpha}}[1-\Lambda_{\rm{\alpha}}\bar{\rho}^2/3]^{\frac{1}{2}}} d(\cos x)\sin^2 x \right|
    \\
    &\qquad \qquad \qquad \qquad=
    \left(\frac{3}{|\Lambda_{\rm{\alpha}}|}\right)^2 \frac{A_3(\bar{\rho})}{\bar{\rho}^3}
    \left|-\frac{2}{3}+\sigma_{\rm{\alpha}}\left(1-\frac{\Lambda_{\rm{\alpha}}\bar{\rho}^2}{3}\right)^{\frac{1}{2}}-\frac{\sigma_{\rm{\alpha}}^3}{3}\left(1-\frac{\Lambda_{\rm{\alpha}}\bar{\rho}^2}{3}\right)^{\frac{3}{2}}\right|,     \label{V4}
\end{align}
where $\rm{\alpha}=\rm{i}$ or $\rm{o}$, and we use coordinates in which the metric is
\begin{align}
    ds^2=\frac{3}{\Lambda_{\rm{i}}}dx^2+\frac{3}{\Lambda_{\rm{i}}}\sin^2 x \,d \Omega_3.
\end{align}
Here, $d \Omega_3$ is the metric for the unit 3-sphere.
In the case of $\Lambda_{\rm{i}}<0$ and  $\sigma_{\rm{i}}=+1$, $V_4(\bar{\rho},\sigma_{\rm{i}},\Lambda_{\rm{i}}(<0))$ is obtained by replacing trigonometric function with hyperbolic functions in \eqref{V4}.

\section{Constraints of the parameter from the Swampland Conjecture and stability}
\label{main}
This section is the main part of the paper. The goal of this section is to bound the parameters using the Swampland Conjectures and the vacuum stability condition.
Here, the parameters refer to the magnetic flux quantization number $M$, the string scale ${{\cal{M}}_{\rm{string}}}$, and the compactification energy scale  ${{\cal{M}}_{\rm{compact}}}$.
Since the scale of the domain wall is approximately on the string scale,\footnote{For instance, the tensions of D-brane and NS5-brane are $1/g_s$ and $1/g_s^2$ in the string unit, respectively. Here $g_s$ is the string coupling. Assuming that $g_s$ is not too small, the tension of the brane is of the same order as ${{\cal{M}}_{\rm{string}}}$.} we take ${{\cal{M}}_{\rm{string}}}\sim {{\cal{M}}_{\rm{wall}}}$.
We introduce the parameters $(\alpha,\beta,\gamma)$ as follows:
\begin{align}
 M_{\rm{o}}^2-M_{\rm{i}}^2 =: 10^{\alpha}, \quad
 {{\cal{M}}_{\rm{wall}}} \,[{\rm{GeV}}]=: 10^{\beta}, \quad
 {{\cal{M}}_{\rm{compact}}} \,[{\rm{GeV}}]=: 10^{\gamma}.
\label{eq:alpha_beta_gamma}\end{align}
We will show that the WGC, the TCC, and the stability condition lead to non-trivial constraints on these parameters.

\subsection{Weak Gravity Conjecture}
\label{sub:WGC}
The WGC~\cite{Arkani-Hamed:2006emk} is to postulate that gravity is always the weakest force in any consistent quantum field theory with gravity. When a theory is given with a $p$-form gauge field weakly coupled to Einstein gravity, there must exist an electrically charged state satisfying
\begin{align}
    \gamma m \leq Q
    \label{WGCdefine}
\end{align}
where $m$ is the tension of the membrane, $Q$ is the physical charge, and $\gamma \sim \sqrt{G}$ is the charge to tension ratio of an extremal black brane in the theory. 
In this paper, we assume that the WGC is valid at any energy scale and impose the WGC on the domain wall solutions in the IR~\cite{Ooguri:2016pdq,Bedroya:2020rac}. 

From Eq.~\eqref{action1} and \eqref{eq:E_difference}, we obtain $Q=e=E_{\rm{o}}-E_{\rm{i}}$, and the WGC for a domain wall is
\begin{align}
   \sqrt{G}m&\lesssim E_{\rm{o}}-E_{\rm{i}}.
    \label{WGC1}
\end{align}
From this, the following equation is satisfied
\begin{align}
    (E_{\rm{o}}^2-E_{\rm{i}}^2)-Gm^2\gtrsim 0.
    \label{WGC2}
\end{align}
From Eqs.~\eqref{sigmainandout} and \eqref{WGC2}, we observe $\sigma_{\rm{o}}=\sigma_{\rm{i}}=+1$.
Consequently, for $\Lambda_{\rm{o}}>\Lambda_{\rm{i}}$, the membrane nucleation is always possible from Table~\ref{T1}.

\begin{figure}[t]
    \begin{tikzpicture}
        \node at (0,0) {\includegraphics[width=0.4\textwidth]{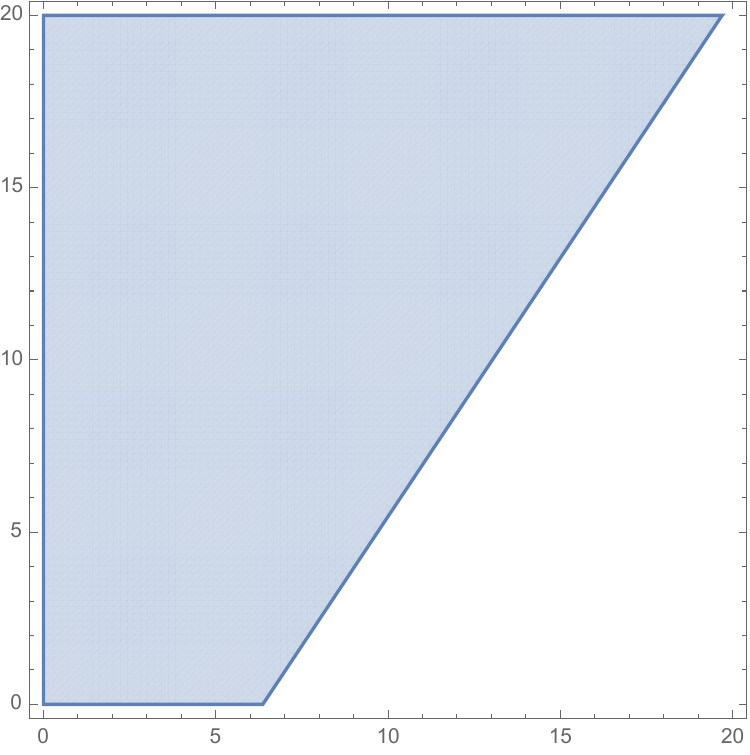}};
        \node at (8,0) {\includegraphics[width=0.4\textwidth]{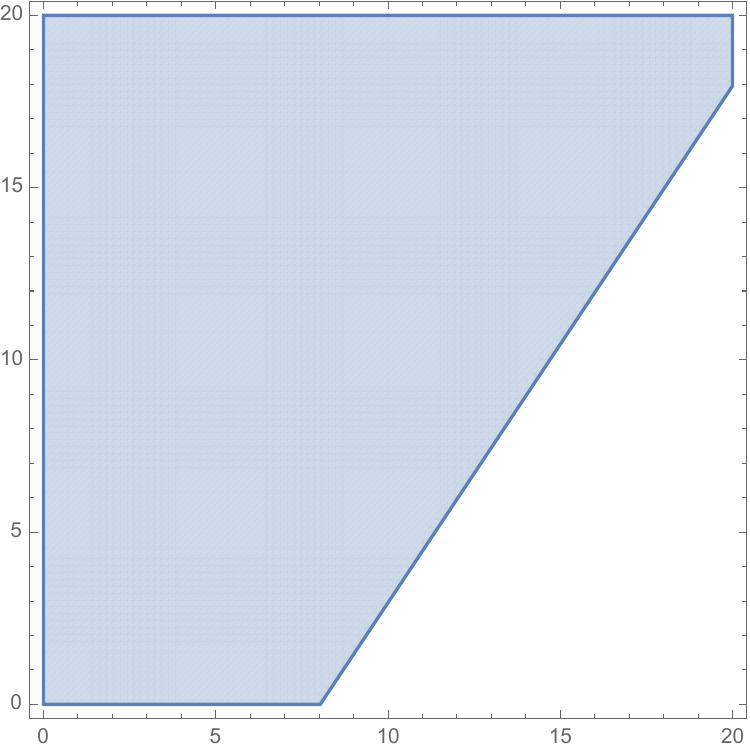}};
        \node[gray!25!blue] at (0,2) {\scalebox{.9}{consistent with WGC}};
        \node[gray!25!black] at (0,-3.5) {\scalebox{.9}{$\beta=\log_{10}\left({{\cal{M}}_{\rm{wall}}}[{\rm{GeV}}]\right)$}};
        \node[gray!25!black] at (-3.5,0) {\rotatebox{90}{\scalebox{.9}{$\gamma=\log_{10}\left({{\cal{M}}_{\rm{compact}}}[{\rm{GeV}}]\right)$}}};
        \node[gray!25!red] at (1.7,-1) {\scalebox{.9}{$M_{\rm{o}}^2-M_{\rm{i}}^2=10^{0}$}};
        \node[gray!25!red] at (10,-1) {\scalebox{.9}{$M_{\rm{o}}^2-M_{\rm{i}}^2=10^{10}$}};
    \end{tikzpicture}
    \caption{Allowed regions by the WGC bound \eqref{WGCbound1} in the $(\beta,\gamma)$ plane for $\alpha=0$ (left) and $\alpha=10$ (right).}
    \label{Fig;WGC}
\end{figure}
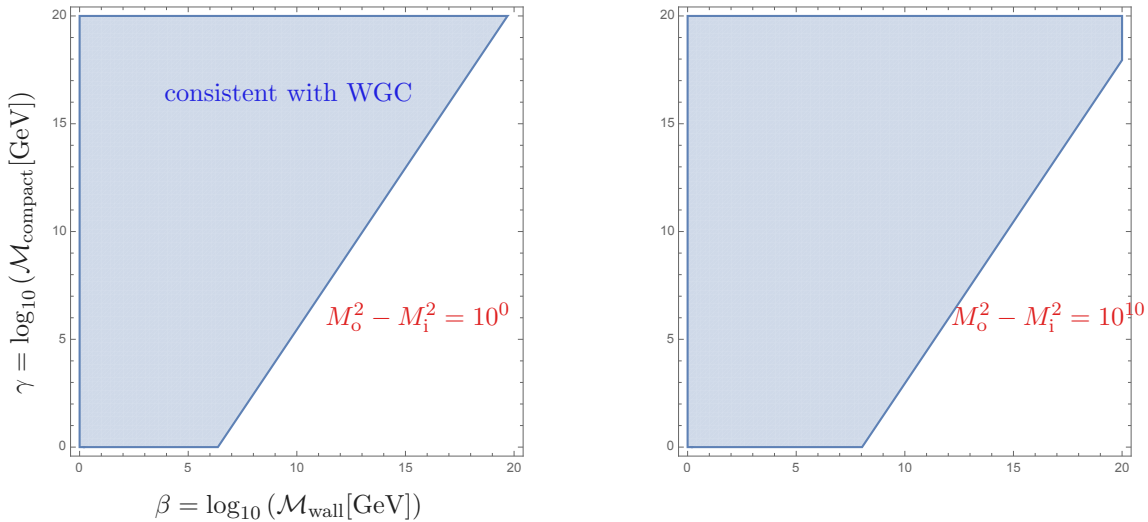

The relationship between $E$ and the flux quantization number $M$ is derived from Eq.\eqref{Ind2d2}
\begin{align}
   E= 2\pi \frac{M}{\cal{A}},
\label{eq:E_and_M}\end{align}
where ${\cal{A}}$ is the volume of the 2d compact manifold ${\mathcal{M}}^2$ and we define ${\cal{M}}_{\rm{compact}}$ as ${\cal{A}}=:({{\cal{M}}_{\rm{compact}}})^{-2}.$ The tension of the membrane is given by
\begin{align}
    m \sim ({{\cal{M}}_{\rm{string}}})^3.
\end{align}
In terms of $\alpha,\,\beta$ and $\gamma$, Eq.\eqref{WGC2} becomes
\begin{align}
    \alpha \geq -39+6\beta-4\gamma,
    \label{WGCbound1}
\end{align}
where we have used $G=6.709 \times  10^{-39}$ [GeV$^{-2}$].
In Figure \ref{Fig;WGC}, we show the allowed regions of the parameters $\beta,\,\gamma$ by the WGC bound \eqref{WGCbound1} for the choice $\alpha=0$. The implicit assumption here is that the membrane with the $\mathcal{O}(1)$ charge satisfies the WGC. Strictly speaking, this is not necessarily the case as the conjecture itself just states there exists one WGC state. However, to the best of our knowledge, we are not aware of an example that parametrically violates our assumption.

\subsection{Stability condition of the Universe}
\label{sub:STB}
We impose the vacuum stability condition on our universe.
Namely, the rate of the membrane creation must be less than the age of the universe: 
\begin{align}
    \frac{\Gamma}{H^4}=\frac{Pe^{-B}}{H^4}\leq 1,
\label{condition1}
\end{align}
where $\Gamma_{{\rm{flat}}\to{\rm{AdS}}}$ is the membrane nucleation rate, and $H$ is the current Hubble parameter. 
As the value of the observed cosmological constant is quite small, we consider the transition from the flat spacetime ($\Lambda_{\rm{o}}=0$) to AdS spacetime ($\Lambda_{\rm{i}}=-4\pi G(E_{\rm{o}}^2-E_{\rm{i}}^2)<0$). 
As we have discussed in the previous subsection, membrane nucleation always occurs thanks to the WGC.

From Eqs.~\eqref{B} and \eqref{eq:rhobar}, the bounce action $B$ and the membrane radius $\bar{\rho}$ are given by (see Appendix~\ref{Bouncecal1} for details)
\begin{align}
    &B\sim \frac{1}{2}\pi^2m\bar{\rho}^3,
    \quad
    \bar{\rho}=\frac{m}{\left[\frac{1}{6}(E_{\rm{o}}^2-E_{\rm{i}}^2)-2\pi G m^2\right]}.
    \label{B2}
\end{align}
The prefactor $P$ is computed as  \cite{Garriga:1993fh,Bedroya:2020rac}
\begin{align}
    P\simeq {m^2 \bar{\rho}^2}=\frac{m^4}{\left[\frac{1}{6}(E_{\rm{o}}^2-E_{\rm{i}}^2)-2\pi G m^2\right]^2} .
    \label{prefactor}
\end{align}
By plugging \eqref{B2} and \eqref{prefactor} into \eqref{condition1}, we get
\begin{align}
   \frac{\Gamma}{H^4}=\frac{m^4}{H^4\left[\frac{1}{6}(E_{\rm{o}}^2-E_{\rm{i}}^2)-2\pi G m^2\right]^2}  
   \exp\left({-\frac{\pi^2 m^4}{2\left[\frac{1}{6}(E_{\rm{o}}^2-E_{\rm{i}}^2)-2\pi G m^2\right]^3}}\right)\leq 1.
   \label{conditionstable}
\end{align}

\begin{figure}[h]
\begin{tikzpicture}
        \node at (0,0) {\includegraphics[width=0.4\textwidth]{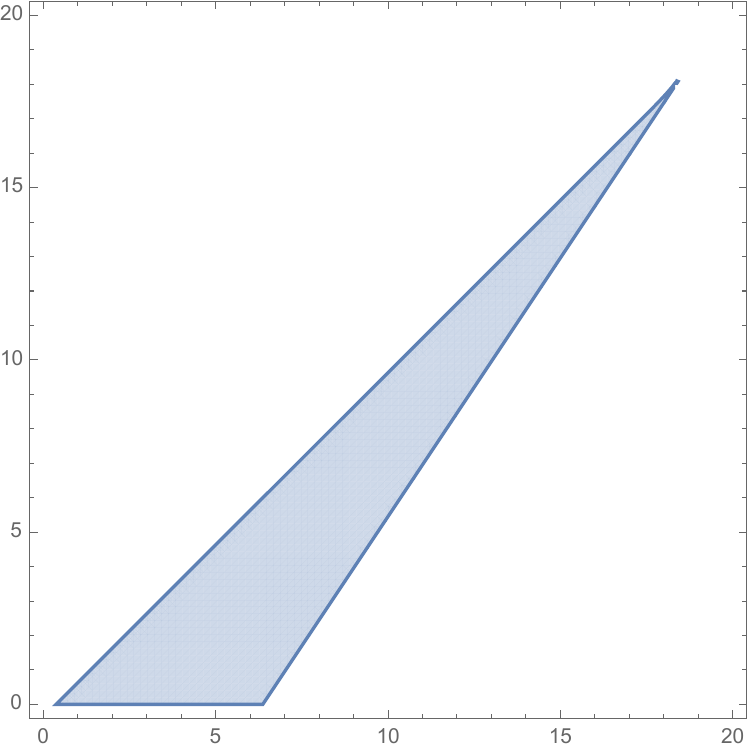}};
        \node at (8,0) {\includegraphics[width=0.4\textwidth]{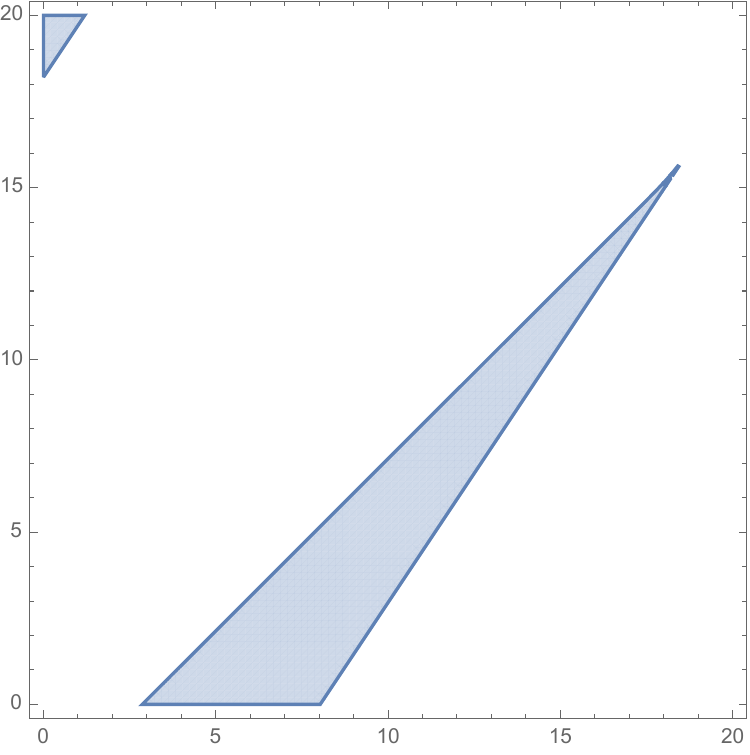}};
        \node[gray!25!blue] at (-1.25,-2) {\scalebox{.9}{Stable}};
        \node[gray!25!black] at (0,-3.5) {\scalebox{.9}{$\beta=\log_{10}\left({{\cal{M}}_{\rm{wall}}}[{\rm{GeV}}]\right)$}};
        \node[gray!25!black] at (-3.5,0) {\rotatebox{90}{\scalebox{.9}{$\gamma=\log_{10}\left({{\cal{M}}_{\rm{compact}}}[{\rm{GeV}}]\right)$}}};
        \node[gray!25!red] at (1.7,-1) {\scalebox{.9}{$M_{\rm{o}}^2-M_{\rm{i}}^2=10^{0}$}};
        \node[gray!25!red] at (10,-1) {\scalebox{.9}{$M_{\rm{o}}^2-M_{\rm{i}}^2=10^{10}$}};
    \end{tikzpicture}
\caption{Allowed regions by Eq.\eqref{conditionstable} in the $(\beta,\gamma)$ plane for $\alpha=0$ (left) and $\alpha=10$ (right).
The island in the upper left region of the right panel corresponds to the situation where $B\ll1$ and $P/H^4<1$. We do not take the island region seriously since it appears below the TeV scale and is excluded by the experiments.}
\label{Fig;STB}
\end{figure}
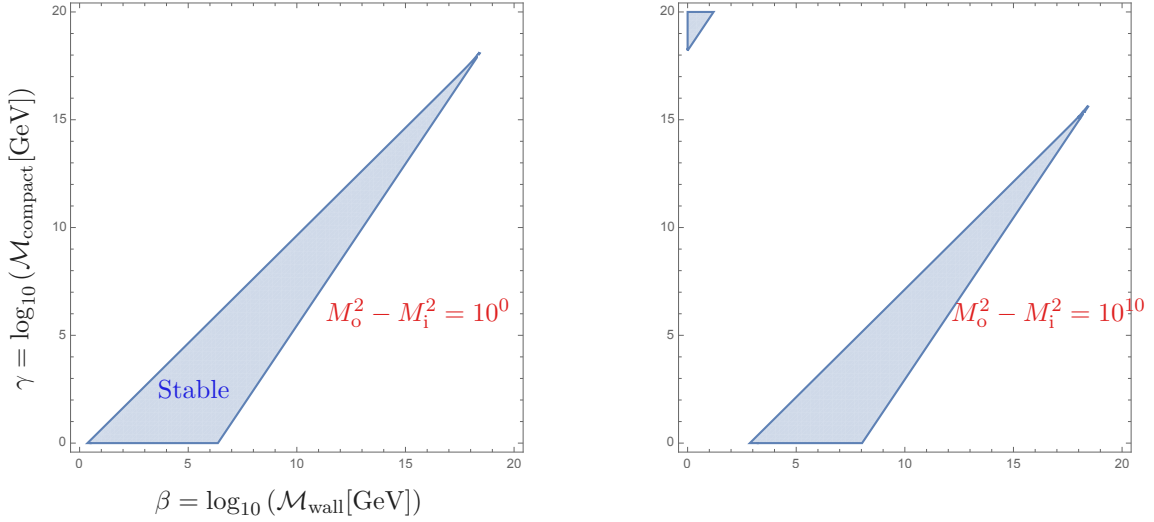

Figure \ref{Fig;STB} shows the constraints on the model parameters coming from Eqs.\eqref{WGCbound1} and \eqref{conditionstable}. 
The allowed region corresponds to the shaded region, which is surrounded by two lines. The upper line corresponds to \eqref{conditionstable} and the lower line comes from the WGC condition \eqref{WGCbound1}.

\subsection{Trans-Planckian Censorship Conjecture}
\label{sub:TCC}

The TCC is one of the Swampland Conjecture, which provides an upper bound on the lifetime of the de Sitter vacua. The TCC states that the expansion of the universe must slow down before all Planckian modes are stretched beyond the Hubble radius~\cite{Bedroya:2019snp,Bedroya:2019tba}. 
When the TCC is broken, the Planckian quantum fluctuations exit the Hubble horizon, freeze out, and become classical. The TCC prohibits such a weird situation.

Although we choose the 2-form flux $F_2$ in such a way that three generations of the chiral fermion are reproduced, the theory admits other vacua with different flux quantization number. In particular, the vacua with $M>3$ has a positive cosmological constant.  Such a vacuum will decay into another vacuum with a smaller $M$.

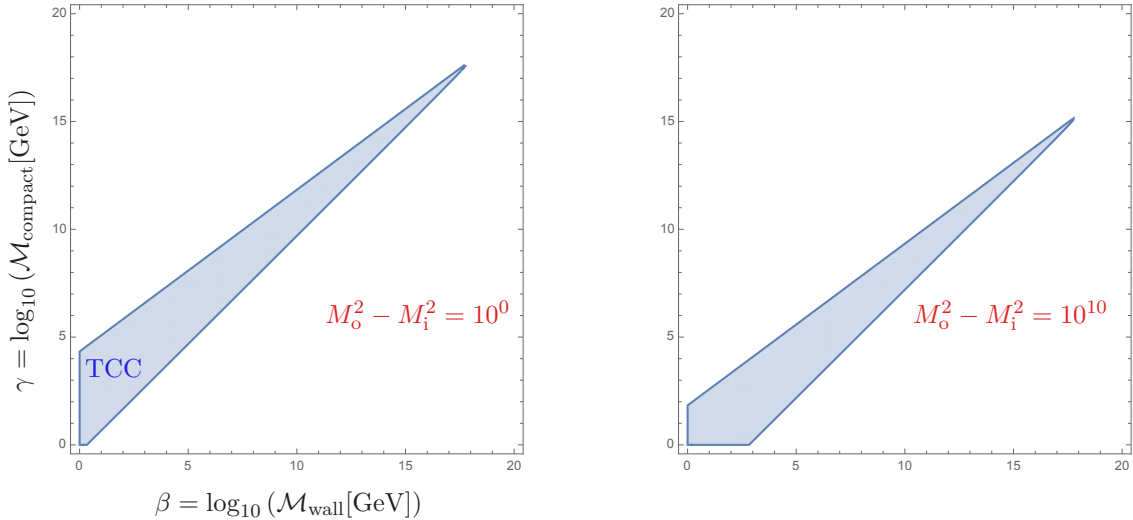
\begin{figure}[h]
\begin{tikzpicture}
        \node at (0,0) {\includegraphics[width=0.4\textwidth]{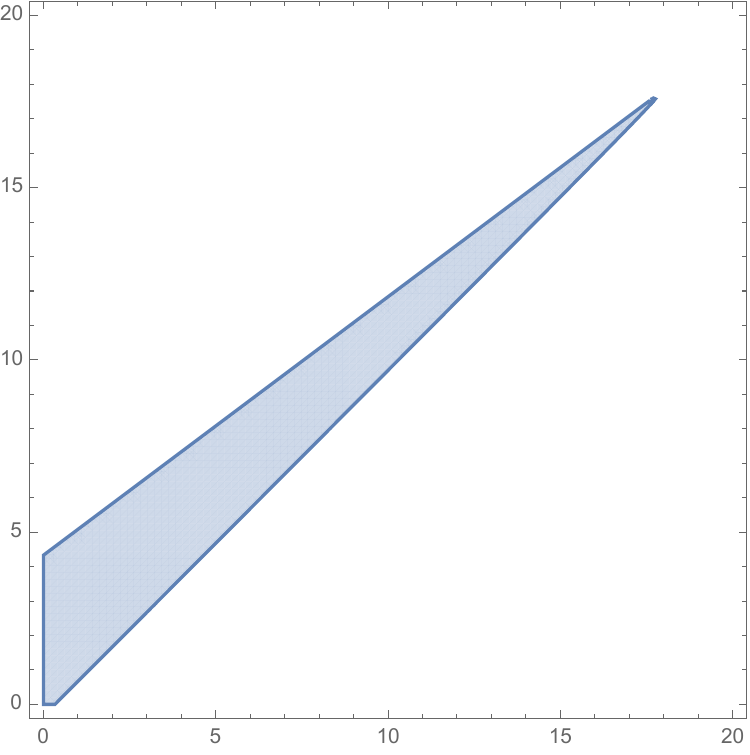}};
        \node at (8,0) {\includegraphics[width=0.4\textwidth]{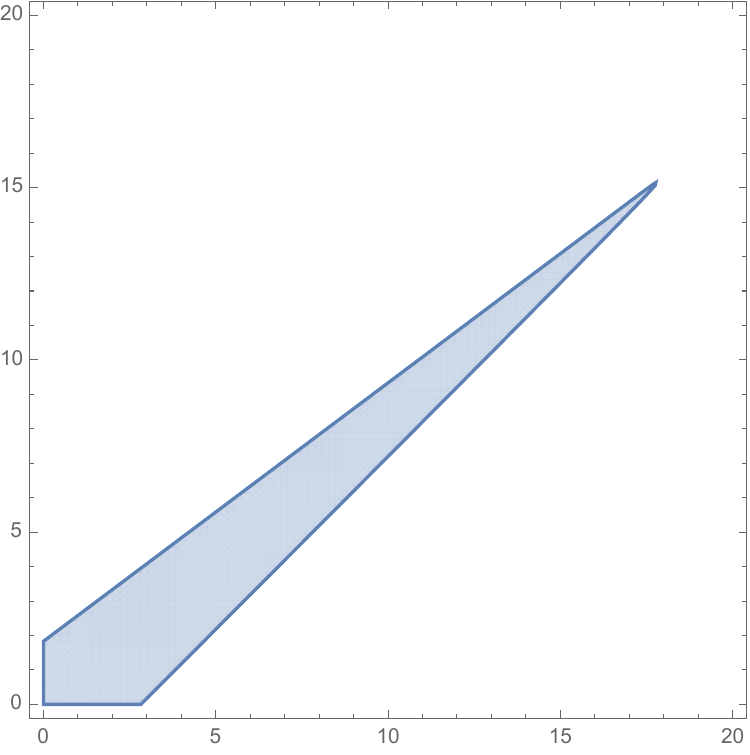}};
        \node[gray!25!blue] at (-2.3,-1.7) {\scalebox{.9}{TCC}};
        \node[gray!25!black] at (0,-3.5) {\scalebox{.9}{$\beta=\log_{10}\left({{\cal{M}}_{\rm{wall}}}[{\rm{GeV}}]\right)$}};
        \node[gray!25!black] at (-3.5,0) {\rotatebox{90}{\scalebox{.9}{$\gamma=\log_{10}\left({{\cal{M}}_{\rm{compact}}}[{\rm{GeV}}]\right)$}}};
        \node[gray!25!red] at (1.7,-1) {\scalebox{.9}{$M_{\rm{o}}^2-M_{\rm{i}}^2=10^{0}$}};
        \node[gray!25!red] at (9.5,-1) {\scalebox{.9}{$M_{\rm{o}}^2-M_{\rm{i}}^2=10^{10}$}};
    \end{tikzpicture}
    \caption{Allowed regions by Eq.\eqref{TCCcondition2} in the $(\beta,\gamma)$ plane for $\alpha=0$ (left) and $\alpha=10$ (right). }
    \label{Fig;TCC}
\end{figure}

Since the Swampland conjectures are applicable to all the vacua in the theory, the decay rate of the de Sitter space is fast enough according to the TCC.\footnote{We assume that the membrane nucleation is the dominant decay process.}
With this in mind, in this subsection, we consider a hypothetical transition from a de Sitter space to a flat space. 
Specifically, we set $E_{\rm{o}}=E_{\rm{dS}}$, $\Lambda_{\rm{o}}=\Lambda_{\rm{dS}}(>0)$, $E_{\rm{i}}=E_{\rm{flat}}$ and $\Lambda_{\rm{i}}=0$. Note that these parameters are related as
\begin{align}
    E_{\rm{flat}}=E_{\rm{dS}}-e, \quad
    \Lambda_{\rm{dS}}=4\pi G( E_{\rm{dS}}^2-E_{\rm{flat}}^2).
\end{align}
The TCC provides the following restrictions on this transition~\cite{Bedroya:2019snp,Bedroya:2020rac}
\begin{align}
\frac{\Gamma_{{\rm{dS}}\to{\rm{flat}}}}{{H^{\prime}}^4}=\frac{P^{\prime}e^{-B^{\prime}}}{{H^{\prime}}^4}>1 ,
    \label{TCCcondition}
\end{align}
where $\Gamma_{{\rm{dS}}\to{\rm{flat}}}$ is the membrane nucleation rate, and $H^{\prime}=\Lambda_{\rm{dS}}^{1/2}$. 
The bounce action $B^\prime$ and the membrane radius $\bar{\rho}$ are given by (see Appendix~\ref{Bouncecal2} for details)
\begin{align}
    &B^{\prime}\sim 
    \frac{1}{2}\pi^2m\bar{\rho}^{{\prime}^3},
    \label{B3}\\
    &\bar{\rho}^{\prime}
    =\left\{\frac{4\pi G}{3}(E_{\rm{dS}}^2-E_{\rm{flat}}^2)
    +\frac{1}{m^2}\left[\frac{1}{6}(E_{\rm{dS}}^2-E_{\rm{flat}}^2)-2\pi G m^2\right]^{2}\right\}^{-\frac{1}{2}}.\label{radius_dS}
\end{align}
The prefactor reads \cite{Bedroya:2020rac}
\begin{align}
    P^{\prime} \simeq m^{\prime2} \bar{\rho}^{\prime2}.
    \label{prefactor2}
\end{align}
By substituting \eqref{B3} and \eqref{prefactor2} into \eqref{TCCcondition}, we get
\begin{align}  \frac{m^{\prime2}\bar{\rho}^{\prime2}}{H^{\prime4}}\exp{\left(-\frac{1}{2}\pi^2m\bar{\rho}^{{\prime}^3}\right)}>1 .
    \label{TCCcondition2}
\end{align}
Figure \ref{Fig;TCC} shows how Eq.\eqref{TCCcondition2} constrains the values of $\beta$ and $\gamma$ for $\alpha=0$ and $\alpha=10$. The allowed region corresponds to the shaded region, and is surrounded by two lines. 
The upper line corresponds to $P^{\prime}>{H^{\prime}}^4$ and the lower line corresponds to $B^{\prime}<{\mathcal{O}}(1)$ to eliminate the exponential suppression. 

\begin{figure}[h]
\begin{tikzpicture}
        \node at (0,0) {\includegraphics[width=0.4\textwidth]{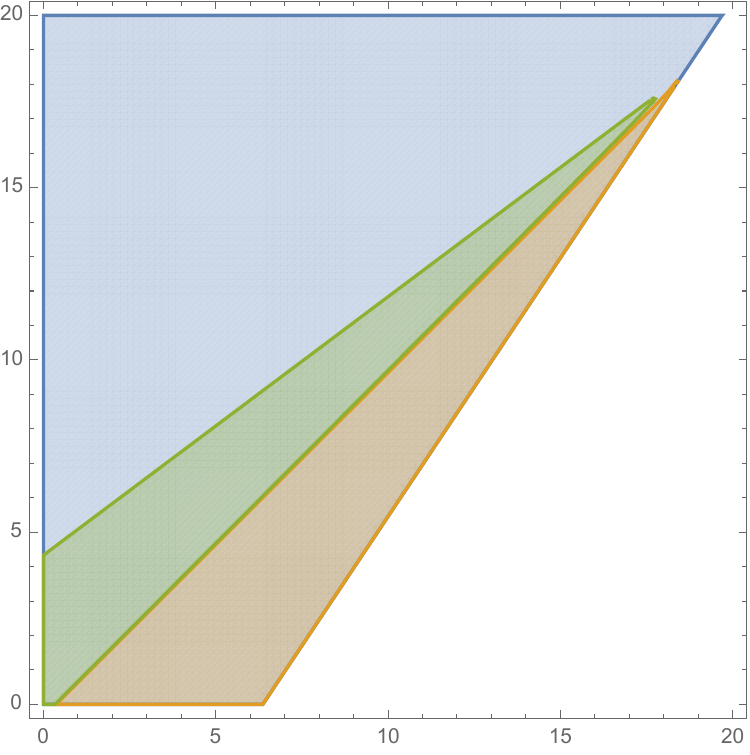}};
        \node at (8,0) {\includegraphics[width=0.4\textwidth]{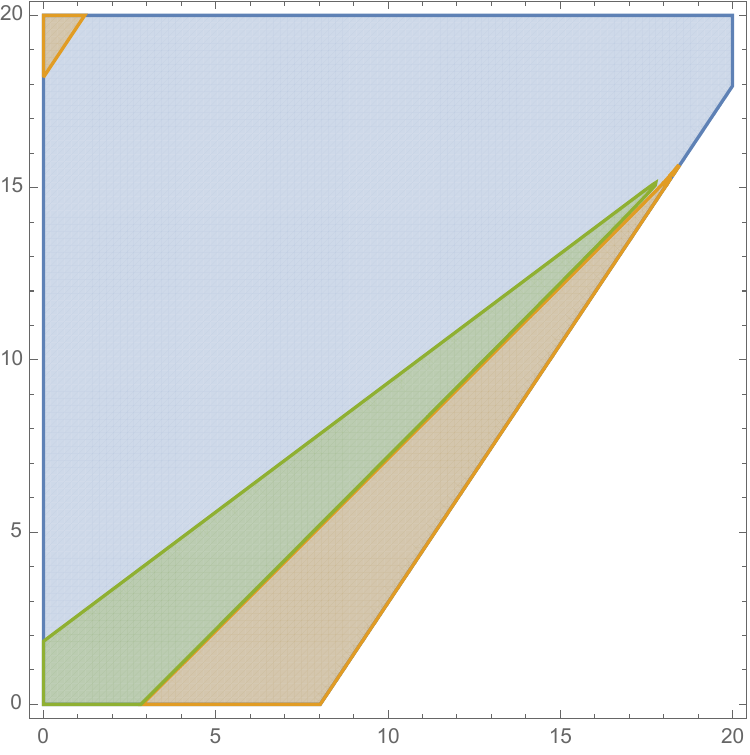}};
        \node[gray!25!blue] at (-0.2,2) {\scalebox{.9}{Consistent with WGC}};
        \node[gray!25!orange] at (-1.25,-2) {\scalebox{.9}{Stable}};
        \node[gray!25!green] at (-2.3,-1.7) {\scalebox{.9}{TCC}};
        \node[gray!25!black] at (0,-3.5) {\scalebox{.9}{$\beta=\log_{10}\left({{\cal{M}}_{\rm{wall}}}[{\rm{GeV}}]\right)$}};
        \node[gray!25!black] at (-3.5,0) {\rotatebox{90}{\scalebox{.9}{$\gamma=\log_{10}\left({{\cal{M}}_{\rm{compact}}}[{\rm{GeV}}]\right)$}}};
        \node[gray!25!red] at (1.7,-1) {\scalebox{.9}{$M_{\rm{o}}^2-M_{\rm{i}}^2=10^{0}$}};
        \node[gray!25!red] at (10,-1) {\scalebox{.9}{$M_{\rm{o}}^2-M_{\rm{i}}^2=10^{10}$}};
    \end{tikzpicture}
    \caption{Allowed regions in the $(\beta,\gamma)$ plane for $\alpha=0$ (left) and $\alpha=10$ (right). The blue region represents the WGC bound \eqref{WGCbound1} and the orange region represents the stability condition \eqref{conditionstable}. The green region corresponds to the TCC bound \eqref{TCCcondition2}.}
    \label{Fig;ALL}
\end{figure}
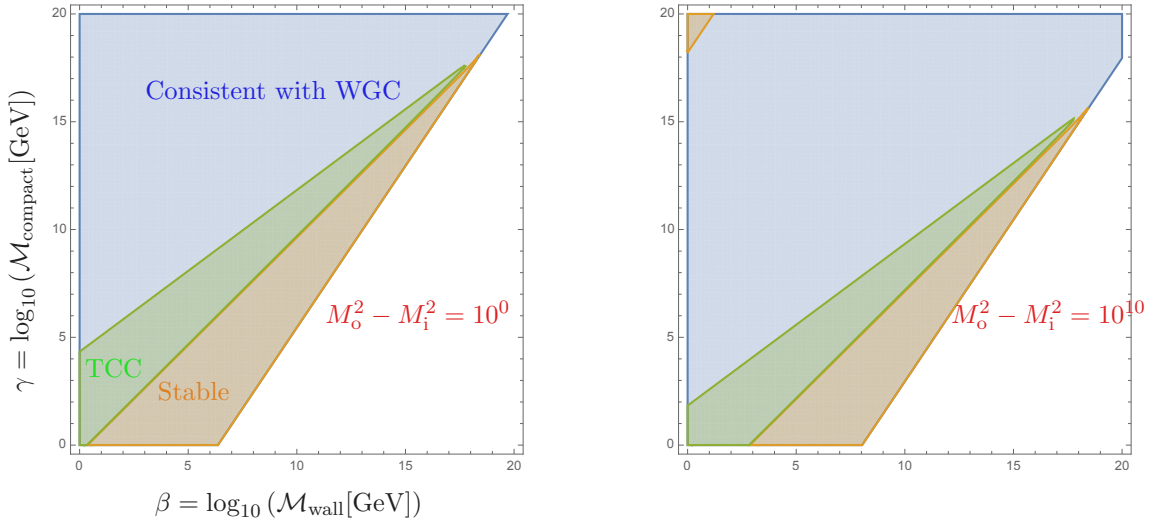

Finally, we summarize the results of the parameter constraints from the WGC, the stability, and the TCC in Figure \ref{Fig;ALL}. As we can see from the figure, there is no parameter space that simultaneously satisfies the stability conditions and the TCC in both the $\alpha=0$ and $10$ cases.

\section{Conclusions}
\label{Conclusion}
In this paper, we have investigated the constraints on the parameters of the magnetized extra-dimensional models: the flux quantization number, the compactification scale, and the string scale. The Atiyah-Singer index theorem for a 2d compact manifold (a two-sphere $S^2$, a 2d torus $T^2$, or a compact hyperbolic manifold $H^2/\Gamma$) with magnetic flux shows that the number of chiral zero modes is determined by the flux quantization number. Therefore, the generation number is equal to the flux quanta. However, the value of the flux quanta can be any integer. 

In order to constrain these parameters of the theory, we have considered some of the Swampland Conjectures. In Section \ref{Membranecreation}, we have reviewed a Brown-Teitelboim picture of membrane creation in the presence of the 4-form antisymmetric tensor field $F_4$. By taking a Hodge dual, the magnetized extra dimension is equivalent to turning on the $F_4$ background on 4d spacetime.

In Section \ref{main}, we have obtained the restrictions on the parameters. First, we have considered the WGC in subsection \ref{sub:WGC}. The WGC gave the condition \eqref{WGCbound1} among the flux quantization number, the compactification scale, and the string scale. From this, the region in the parameter space $(\beta,\gamma)$ is constrained as shown in Figure \ref{Fig;WGC}. Then, we have imposed the stability condition \eqref{conditionstable}.
As shown in Figure \ref{Fig;STB}, the stability condition gives a strong constraint for the parameter space $(\beta,\gamma)$. 
In subsection \ref{sub:TCC}, we have considered the hypothetical transition from a de Sitter space to a flat universe. The constraints on the transition rate provided by the TCC are shown in Figure~\ref{Fig;TCC}.
Figure~\ref{Fig;ALL} summarizes the constraints on the parameter space $(\beta,\gamma)$ imposed by the WGC, the stability conditions, and the TCC. 
Regardless of the value of $\alpha$, no parameter space satisfies both the stability condition and the TCC at the same time.

An interesting future direction is to explore other dimensions and other compactifications.
In particular, the compactifications with singularities (e.g. magnetized $T^2/{\mathbb{Z}}_N$ orbifolds) are interesting because the index is determined not only by the magnetic quantization number but also by the contribution of the singularities~\cite{Sakamoto:2020pev,Sakamoto:2020vdy,Kobayashi:2022tti,Imai:2022bke}. In the case of two dimensions, it was not possible to simultaneously satisfy the stability conditions and the TCC on smooth manifolds ($S^2$, $T^2$, and $H^2/\Gamma$). However, investigating how these conditions are modified in models with singularities is of particular interest.

\section*{Acknowledgments}
This work was supported by MEXT Leading Initiative for Excellent Young Researchers Grant No.JPMXS0320210099 (Y.H.) and JSPS KAKENHI Grants No.24H00976 (Y.H.), 24K07035 (Y.H.).

\appendix

\section{The bounce action}
In this appendix, we show the detailed calculations of $B$.

\subsection{Transition from the flat to AdS spacetime}
\label{Bouncecal1}
From \eqref{B2}, the membrane radius $\bar{\rho}$ is 
\begin{align}
\bar{\rho}&=\frac{m}{\left[\frac{1}{6}(E_{\rm{o}}^2-E_{\rm{i}}^2)-2\pi G m^2\right]} .
\end{align}
When evaluating the dimensionless quantity $\bar{\rho}^2\Lambda_i$ here
\begin{align}
    \bar{\rho}^2\Lambda_i
    &=\left(\frac{m}{\left[\frac{1}{6}(E_{\rm{o}}^2-E_{\rm{i}}^2)-2\pi G m^2\right]}\right)^2 \left(-4\pi G(E_{\rm{o}}^2-E_{\rm{i}}^2)\right).
\end{align}
In the case of $\frac{\bar{\rho}^2|\Lambda_i|}{3}\ll1$, 
the bounce action can be approximated as follows;
\begin{align}
B&=mA_3(\bar{\rho})+\left\{\left[-\frac{2\Lambda_{\rm{i}}}{16\pi G}V_4(\bar{\rho},\sigma_{\rm{i}},\Lambda_{\rm{i}})-\frac{3\sigma_{\rm{i}}}{8\pi G}\left(\bar{\rho}^{-2}-\frac{2\Lambda_{\rm{i}}}{6}\right)^{\frac{1}{2}}A_3(\bar{\rho})\right]-({\rm{i}}\to {\rm{o}})\right\}
\\
&=2\pi^2m\bar{\rho}^3-\frac{3\pi}{4G}\bar{\rho}^2
\left\{-1+\left(1-\frac{\bar{\rho}^2\Lambda_i}{3}\right)^{\frac{1}{2}}+\frac{3}{\bar{\rho}^2\Lambda_i}\left|\frac{2}{3}-\left(1-\frac{\bar{\rho}^2\Lambda_i}{3}\right)^{\frac{1}{2}}+\frac{1}{3}\left(1-\frac{\bar{\rho}^2\Lambda_i}{3}\right)^{\frac{3}{2}}\right|\right\}\\
&\simeq 2\pi^2m\bar{\rho}^3+\frac{\pi \bar{\rho}^4\Lambda_{\rm{i}}}{16G}+\frac{\pi \bar{\rho}^6\Lambda_{\rm{i}}^2}{288G}+\cdots\\
&= 2\pi^2m\bar{\rho}^3 \left\{1+\frac{\bar{\rho}\Lambda_{\rm{i}}}{32\pi Gm}\left(1+\frac{\bar{\rho}^2\Lambda_{\rm{i}}}{18}+\cdots\right)\right\}\\
&= 2\pi^2m\bar{\rho}^3 \left\{1-\frac{3}{4}\left(1-\frac{2\pi G m^2}{\frac{1}{6}(E_{\rm{o}}^2-E_{\rm{i}}^2)}\right)^{-1}\left(1+\frac{\bar{\rho}^2\Lambda_{\rm{i}}}{18}+\cdots\right)\right\}\\
&\simeq \frac{1}{2}\pi^2m\bar{\rho}^3.
\label{STBsimple}
\end{align}
In the third equality, we have assumed $\frac{\bar{\rho}^2|\Lambda_i|}{3}\ll1$. Additionally, in the last equality, we have used the WGC condition.

\begin{figure}[h]
\begin{tikzpicture}
        \node at (0,0) {\includegraphics[width=0.4\textwidth]{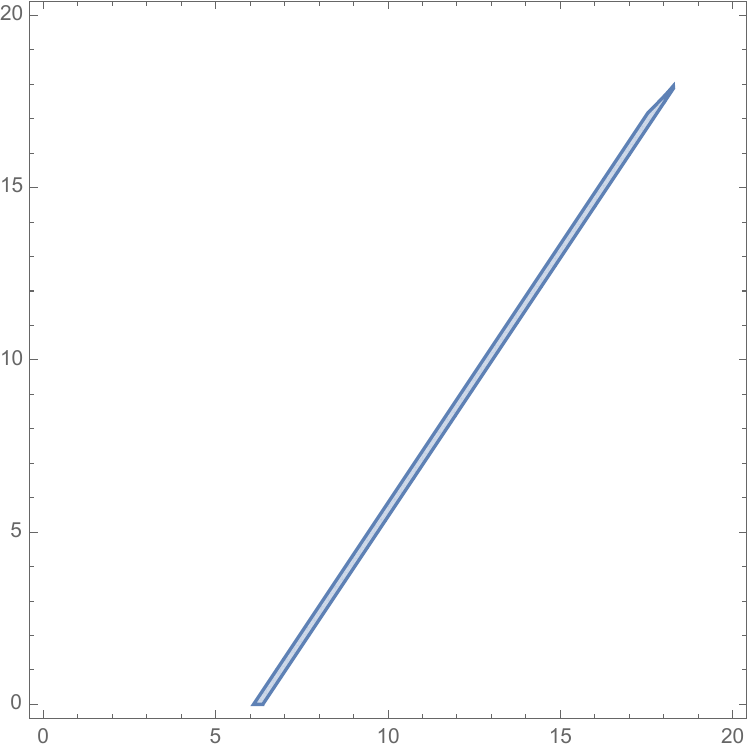}};
        \node at (8,0) {\includegraphics[width=0.4\textwidth]{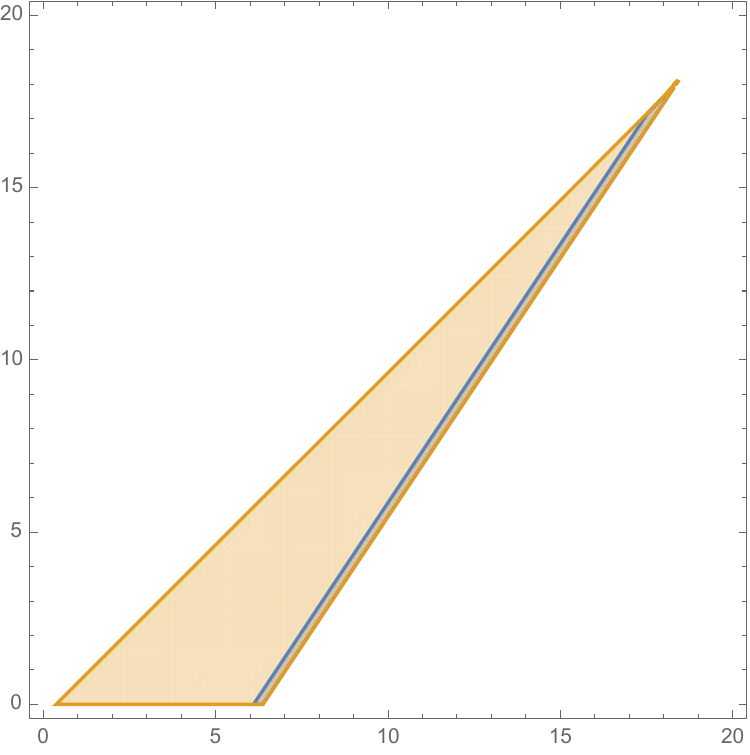}};
         \node[gray!25!black] at (0,-3.5) {\scalebox{.9}{$\beta=\log_{10}\left({{\cal{M}}_{\rm{wall}}}[{\rm{GeV}}]\right)$}};
         \node[gray!25!black] at (-3.5,0) {\rotatebox{90}{\scalebox{.9}{$\gamma=\log_{10}\left({{\cal{M}}_{\rm{compact}}}[{\rm{GeV}}]\right)$}}};
    \end{tikzpicture}
     \caption{
     The left panel illustrates the region where the stability condition calculated by the bounce action \eqref{STBfull} are satisfied when $\alpha=0$, in the region where the approximation is not applicable ($\frac{\bar{\rho}^2|\Lambda_i|}{3}\sim 1$ or $\frac{\bar{\rho}^2|\Lambda_i|}{3}>1$).
     In the right panel, the orange region represents the stability condition calculated by the bounce action \eqref{STBsimple}. From this, the region allowed by Eq.\eqref{STBfull} in the region where the approximation is not applicable is encompassed within the allowable region defined by Eq.\eqref{STBsimple}.} 
    \label{Fig;STBfull}
\end{figure}

On the other hand, for the case where $\frac{\bar{\rho}^2|\Lambda_i|}{3}\sim 1$ or $\frac{\bar{\rho}^2|\Lambda_i|}{3}>1$, the bounce action is given by
\begin{align}
B&=mA_3(\bar{\rho})+\left\{\left[-\frac{2\Lambda_{\rm{i}}}{16\pi G}V_4(\bar{\rho},\sigma_{\rm{i}},\Lambda_{\rm{i}})-\frac{3\sigma_{\rm{i}}}{8\pi G}\left(\bar{\rho}^{-2}-\frac{2\Lambda_{\rm{i}}}{6}\right)^{\frac{1}{2}}A_3(\bar{\rho})\right]-({\rm{i}}\to {\rm{o}})\right\}
\\
&=2\pi^2m\bar{\rho}^3-\frac{3\pi}{4G}\bar{\rho}^2
\left\{-1+\left(1-\frac{\bar{\rho}^2\Lambda_i}{3}\right)^{\frac{1}{2}}+\frac{3}{\bar{\rho}^2\Lambda_i}\left|\frac{2}{3}-\left(1-\frac{\bar{\rho}^2\Lambda_i}{3}\right)^{\frac{1}{2}}+\frac{1}{3}\left(1-\frac{\bar{\rho}^2\Lambda_i}{3}\right)^{\frac{3}{2}}\right|\right\}.
\label{STBfull}
\end{align}
In this case, since approximation is not possible, we must evaluate \eqref{STBfull}. The results are shown in Figure \ref{Fig;STBfull}.
From this, the region allowed by Eq.\eqref{STBfull} in the region where the approximation is not applicable is encompassed within the allowable region defined by Eq.\eqref{STBsimple}.
Consequently, we can use \eqref{STBsimple} for the whole parameter region as far as the stability condition is concerned.

\subsection{Transition from the dS to flat spacetime}
\label{Bouncecal2}
From \eqref{radius_dS}, the membrane radius $\bar{\rho}$ is 
\begin{align}
\bar{\rho}^{\prime}
    &=\left\{\frac{4\pi G}{3}(E_{\rm{dS}}^2-E_{\rm{flat}}^2)
    +\frac{1}{m^2}\left[\frac{1}{6}(E_{\rm{dS}}^2-E_{\rm{flat}}^2)-2\pi G m^2\right]^{2}\right\}^{-\frac{1}{2}} .
\end{align}
When evaluating the dimensionless quantity $\bar{\rho}^{{\prime}^2}\Lambda_{\rm{o}}$ here
\begin{align}
    \frac{\bar{\rho}^{{\prime}^2}\Lambda_{\rm{o}}}{3}
    =\frac{4\pi G(E_{\rm{dS}}^2-E_{\rm{flat}}^2)}{3\left\{\frac{4\pi G}{3}(E_{\rm{dS}}^2-E_{\rm{flat}}^2)
    +\frac{1}{m^2}\left[\frac{1}{6}(E_{\rm{dS}}^2-E_{\rm{flat}}^2)-2\pi G m^2\right]^{2}\right\}}
    <1.
\end{align}
Since the second term in the denominator is always positive, the condition $\frac{\bar{\rho}^{{\prime}^2}\Lambda_{\rm{o}}}{3}< 1$ is always satisfied.
In the case of $\frac{\bar{\rho}^{{\prime}^2}\Lambda_{\rm{o}}}{3}\ll1$,
the bounce action can be approximated as follows;
\begin{align}
B&=mA_3(\bar{\rho}^{\prime})+\left\{\left[-\frac{2\Lambda_{\rm{i}}}{16\pi G}V_4(\bar{\rho}^{\prime},\sigma_{\rm{i}},\Lambda_{\rm{i}})-\frac{3\sigma_{\rm{i}}}{8\pi G}\left(\bar{\rho}^{{\prime}^{-2}}-\frac{2\Lambda_{\rm{i}}}{6}\right)^{\frac{1}{2}}A_3(\bar{\rho}^{\prime})\right]-({\rm{i}}\to {\rm{o}})\right\}
\\
&=2\pi^2m\bar{\rho}^{{\prime}^3}-\frac{3\pi}{4G}\bar{\rho}^{{\prime}^2}
 \left\{1-\left(1-\frac{\bar{\rho}^{{\prime}^2}\Lambda_{\rm{o}}}{3}\right)^{\frac{1}{2}}-\frac{3}{\bar{\rho}^{{\prime}^2}\Lambda_{\rm{o}}}\left|-\frac{2}{3}+\left(1-\frac{\bar{\rho}^{{\prime}^2}\Lambda_{\rm{o}}}{3}\right)^{\frac{1}{2}}-\frac{1}{3}\left(1-\frac{\bar{\rho}^{{\prime}^2}\Lambda_{\rm{o}}}{3}\right)^{\frac{3}{2}}\right|\right\} \\
&\simeq 2\pi^2m\bar{\rho}^{{\prime}^3}-\frac{\pi \bar{\rho}^{{\prime}^4}\Lambda_{\rm{o}}}{16G}-\frac{\pi \bar{\rho}^{{\prime}^6}\Lambda_{\rm{o}}^2}{288G}+\cdots\\
&= 2\pi^2m\bar{\rho}^{{\prime}^3} \left\{1-\frac{\bar{\rho}^{\prime}\Lambda_{\rm{o}}}{32\pi Gm}\left(1+\frac{\bar{\rho}^{{\prime}^2}\Lambda_{\rm{o}}}{18}+\cdots\right)\right\}\\
&= 2\pi^2m\bar{\rho}^{{\prime}^3} \left\{1-\frac{3}{4}\left(1+\frac{2\pi G m^2}{\frac{1}{6}(E_{\rm{dS}}^2-E_{\rm{flat}}^2)}\right)^{-1}\left(1+\frac{\bar{\rho}^{{\prime}^2}\Lambda_{\rm{o}}}{18}+\cdots\right)\right\}\\
&\simeq \frac{1}{2}\pi^2m\bar{\rho}^{{\prime}^3}.
\end{align}
In the third equality, we have assumed  $\frac{\bar{\rho}^{{\prime}^2}\Lambda_{\rm{o}}}{3}\ll1$. Additionally, in the last equality, we have used the WGC condition.

On the other hand, for $\frac{\bar{\rho}^{{\prime}^2}\Lambda_{\rm{o}}}{3} \sim 1$, the bounce action is given by
\begin{align}
B&=mA_3(\bar{\rho}^{\prime})+\left\{\left[-\frac{2\Lambda_{\rm{i}}}{16\pi G}V_4(\bar{\rho}^{\prime},\sigma_{\rm{i}},\Lambda_{\rm{i}})-\frac{3\sigma_{\rm{i}}}{8\pi G}\left(\bar{\rho}^{{\prime}^{-2}}-\frac{2\Lambda_{\rm{i}}}{6}\right)^{\frac{1}{2}}A_3(\bar{\rho}^{\prime})\right]-({\rm{i}}\to {\rm{o}})\right\}
\\
&=2\pi^2m\bar{\rho}^{{\prime}^3}-\frac{3\pi}{4G}\bar{\rho}^{{\prime}^2}
 \left\{1-\left(1-\frac{\bar{\rho}^{{\prime}^2}\Lambda_{\rm{o}}}{3}\right)^{\frac{1}{2}}-\frac{3}{\bar{\rho}^{{\prime}^2}\Lambda_{\rm{o}}}\left|-\frac{2}{3}+\left(1-\frac{\bar{\rho}^{{\prime}^2}\Lambda_{\rm{o}}}{3}\right)^{\frac{1}{2}}-\frac{1}{3}\left(1-\frac{\bar{\rho}^{{\prime}^2}\Lambda_{\rm{o}}}{3}\right)^{\frac{3}{2}}\right|\right\}
 \label{appTCCfull}
\end{align}
In this case, since the approximation is not possible, we must evaluate \eqref{appTCCfull}. 
It turns out that no region satisfies the TCC condition.

\bibliography{reference}
\bibliographystyle{TitleAndArxiv}

\end{document}